\documentstyle[preprint,aps,epsf]{revtex}
\begin{document}
\tightenlines
\draft
\newcommand{\be}{\begin{equation}}
\newcommand{\ee}{\end{equation}}
\newcommand{\bea}{\begin{eqnarray}}
\newcommand{\eea}{\end{eqnarray}}
\title{Yang-Lee zeros of the $Q$-state Potts model
in the complex magnetic-field plane}
\author{Seung-Yeon Kim\footnote{Electronic address: kim@cosmos.psc.sc.edu}
and Richard J. Creswick\footnote{Electronic address: creswick.rj@sc.edu}}
\address{Department of Physics and Astronomy,
University of South Carolina, \\
Columbia, South Carolina 29208}
\maketitle
\begin{abstract}
The microcanonical transfer matrix is used to study the
distribution of Yang-Lee zeros of the $Q$-state Potts model
in the complex magnetic-field ($x=e^{\beta h}$) plane
for the first time.
Finite size scaling suggests that at (and below) the critical 
temperature the zeros lie close to, but not on, the unit circle
with the two exceptions of the critical point $x=1$ ($h=0$) itself
and the zeros in the limit $T=0$. 
\end{abstract}

The $Q$-state Potts model\cite{po,wu} in two dimensions is very fertile
ground for the analytical and numerical investigation
of first- and second-order phase transitions.
With the exception of the $Q=2$ Potts (Ising) model
in the absence of a magnetic field\cite{on},
exact solutions for arbitrary $Q$ are not known.
However, some exact results have been established for the $Q$-state
Potts model. For $Q=2$, 3 and 4 there is a second-order
phase transition, while for $Q>4$ the transition is first order\cite{b1}.
From the duality relation the critical temperature is 
known to be $T_c/J=1/ln(1+\sqrt{Q})$\cite{po}.
For $Q=3$ and 4 the critical exponents\cite{ex} are known,
while for $Q>4$ the latent heat\cite{b1}, spontaneous magnetization\cite{b2},
and correlation length\cite{bw} at $T_c$ are also known. 

By introducing the concept of the zeros in the {\it complex}
magnetic-field plane of the grand partition function (Yang-Lee zeros),
Yang and Lee\cite{yl} proposed a mechanism
for the occurrence of phase transitions
in the thermodynamic limit and gained a new insight into
the unsolved problem of the Ising model 
in an arbitrary nonzero external magnetic field. 
They have shown that the distribution
of the zeros of a model determines its critical
behavior. Lee and Yang\cite{ly} also formulated
the celebrated circle theorem which states that
the zeros of the grand partition function 
of the Ising ferromagnet in the complex magnetic-field plane
lie on the unit circle.
In 1964 Fisher\cite{fi} initiated the study of the partition
function zeros in the complex temperature plane (Fisher zeros) for
the square lattice Ising model and  
since that time this problem
has attracted continuous interest.
In particular,
the Fisher zeros of the $Q$-state Potts model in the absence of
a magnetic field have been studied extensively\cite{ma,hu,ba,ck2}.
By numerical methods it has been shown\cite{hu} that for self-dual 
boundary conditions the Fisher zeros of the $Q$-state Potts model
on a finite square lattice are located on the unit circle
in the complex $p$-plane for $Re(p)>0$,
where $p=(y^{-1}-1)/\sqrt{Q}$ and $y=e^{-\beta}$.
The study of the Yang-Lee zeros of the Ising model 
has a long history, and some results have been reported 
in one\cite{ly}, two\cite{ly,ck1}, three\cite{su}, and four\cite{ke} dimensions.
However, except for the one-dimensional Potts model\cite{mg},
the Yang-Lee zeros of the $Q > 2$ Potts models
have never been studied.
In this paper we discuss the Yang-Lee zeros of the $Q$-state Potts
model in two dimensions.

We use an {\it exact} numerical technique for evaluation
of grand partition functions, the microcanonical
transfer matrix ($\mu$TM)\cite{sc,ck1,ck2}.
The bond-energy for the $Q$-state Potts model is (in dimensionless units)
\be
E=\sum_{<i,j>}(1-\delta(\sigma_i,\sigma_j)),
\ee 
where $<i,j>$ indicates a sum over nearest-neighbor pairs,
$\sigma_i=0,...,Q-1$, and $E$ is a positive integer $0\le E\le N_b$
where $N_b$ is the number of bonds on the lattice.
We study the grand partition function
of the Potts model in an external field
which couples to the order parameter
\be
M_q=\sum_{k}\delta(\sigma_k,q),
\ee
where $q$ is a fixed integer between 0 and $Q-1$. 
Note that $0\le M_q\le N_s$ is also an integer and $N_s$ is the 
number of sites on the lattice. 
By $\mu$TM it is possible to obtain {\it exact} integer values
for the number of states with fixed energy $E$ and
fixed order parameter $M$, $\Omega_Q(M,E)$. The grand
partition function in a magnetic field, $h$, is then
a polynomial given by
\be
Z_Q(x,y)=\sum_{M=0}^{N_s}\sum_{E=0}^{N_b}\Omega_Q(M,E)x^M y^E,
\ee
where $x=e^{\beta h}$ and $y=e^{-\beta}$.
We have calculated the grand partition function of the $Q$-state Potts model
on finite $L\times L$ square lattices with self-dual boundary 
conditions\cite{hu} and cylindrical boundary conditions for $3\le Q\le8$.

Figure 1 shows the Yang-Lee zeros
for the three-state Potts model in the complex $x$ plane
at the critical temperature $y_c=1/(1+\sqrt{3})=0.366...$
for $L=4$ and $L=10$ with cylindrical boundary conditions.
Note that unlike the Ising model, the zeros of the three-state
Potts model lie close to, but not on, the unit circle.
The zero farthest from the unit circle is in the neighborhood
of $Arg(x)=\pi$, while the zero closest to the positive 
real axis lies closest to the unit circle.
Note that the zeros for $L=10$ lie on a locus interior to
that for $L=4$. We observe similar behavior for larger values of $Q$.
We expect that in the thermodynamic limit the locus of zeros cuts
the real $x$ axis at the point $x=1$ corresponding to $h=0$.
Table I shows the distance from the origin and the imaginary 
part of the first two zeros of the three-state Potts model
for $3\le L\le 12$. By using the Bulirsch-Stoer (BST) algorithm\cite{bh},
we extrapolated our results for finite lattices to infinite size.
The error bars are twice the difference 
between the $(n-1,1)$ and $(n-1,2)$ approximants.
As one can see, these zeros converge to the critical point, $x=1$,
as described by Yang and Lee\cite{yl,ly}. 

While we lack the circle theorem of Lee and Yang to tell us the location
of the zeros, something can be said about their general behavior
as a function of temperature.
At zero temperature ($y=0$) from Eq. (3) the grand partition
function is
\bea
Z_Q(x,0)&=&\sum_{M}\Omega_Q(M,0)x^M \cr
&=&(Q-1)+x^{N_s}.
\eea
Therefore, the Yang-Lee zeros at $T=0$ are given by
\be
x_k=(Q-1)^{1/N_s}exp[i(2k-1)\pi /N_s],
\ee
where $k=1,...,N_s$.
The zeros at $T=0$ are uniformly distributed on the
circle with radius $(Q-1)^{1/N_s}$
which approaches unity in the thermodynamic limit, independent of $Q$.
At infinite temperature ($y=1$) Eq. (3) becomes
\be
Z_Q(x,1)=\sum_{M=0}^{N_s}\sum_{E=0}^{N_b}\Omega_Q(M,E)x^M.$$
\ee
Because $\sum_{E}\Omega_Q(M,E)$ is simply
${N_s\choose M}(Q-1)^{N_s-M}$,
at $T=\infty$, the grand partition function is given by
\be
Z_Q(x,1)=(Q-1+x)^{N_s},
\ee
and its zeros are $N_s$-degenerate at $x=1-Q$,
independent of lattice size. 
Figure 2 shows the zeros for the three-state Potts model 
at several temperatures with cylindrical boundary conditions.
At $y=0.5y_c$ the zeros are uniformly distributed close to the unit 
circle. As the temperature is increased the edge singularity 
moves away from the real axis and the zeros detach  from the unit circle.
Finally, as $y$ approaches unity, the zeros converge on the point $x=-2$. 

For self-dual boundary conditions \cite{hu}
we observe the same behaviors as those in Figures 1 and 2
for cylindrical boundary conditions. 
$N_s=L^2$ and $N_b=2L^2-L$ for cylindrical boundary conditions,
while $N_s=L^2+1$ and $N_b=2L^2$ for self-dual boundary conditions\cite{hu}.
One of the main differences between two boundary conditions is the number
of zeros, which is equal to $N_s$.
Figure 3 shows the Yang-Lee zeros of the three-state Potts model
at $y=y_c$ for $L=7$ with self-dual and cylindrical boundary conditions.
The difference in the number of zeros between two boundary conditions
results in the difference in the locations of zeros near $x=-1$.
However, as $x$ approaches 1, the zeros for the two different boundary 
conditions are nearly identical. 
We observe that the effect of the boundary condition
on the location of the Yang-Lee zeros near the critical point 
of the Potts model is very small, and in the rest of this paper 
we will consider only cylindrical boundary conditions.

It is clear that the Yang-Lee zeros of the $Q$-state
Potts model do not lie on the unit circle for $Q>2$ 
for any value of $y$ and any finite
value of $L$. However, there is some cause to speculate that for $y\le y_c$
the zeros {\it do} lie on the unit circle in the thermodynamic limit.
Since the zero in the neighborhood of $Arg(x)=\pi$ is always
the farthest from the unit circle, if this zero can be shown to
approach $|x(\pi)|=1$ in the limit $L\to\infty$, all the zeros 
should lie on the unit circle in this limit. In Figure 4 we show
values for $|x(\pi)|$ extrapolated to infinite size using the BST 
algorithm\cite{bh} for $3\le Q\le8$ at $y=0.5y_c$ and $y=y_c$.
From these results it is clear that while the locus of zeros lies 
{\it close} to the unit circle at $y=y_c$, it does not coincide
with it except at the critical point $x=1$.  

Figure 5 shows the BST estimate of the 
modulus of the locus of zeros as a function of angle for 
the three-state Potts model at $y=0.5y_c$, $y=y_c$ and $y=1.2y_c$.
To calculate the extrapolated values for each angle, $\theta$,
we selected the zero whose arguments were closest to $\theta$
for lattices of size $3 \le L \le 12$
for $\theta=0.0, 0.5,...,2.5$ and $\pi$. 
The BST algorithm was then used to extrapolate these values
for finite lattices to infinite size.
The large variation in the size of the error bars is due to
the fact that for a given $\theta$ there may be no zero 
{\it close} to $\theta$ for the smaller lattices.  
In Figure 5 at $y=y_c$ the first four angles are shifted slightly
from the original values ($\theta=0.0$, 0.5, 1.0 and 1.5)
to be distinguished from the results at $y=0.5y_c$.
For $y=0.5y_c$ and $y=y_c$ the first zeros definitely lie on the
point $r(\theta=0)=1$ in the thermodynamic limit.
However, for $y=1.2y_c$ the BST estimates of the modulus and angle of the
first zero are 1.054(2) and 0.09(6).
Therefore, at $y=1.2y_c$ the locus of zeros does not cut the
positive real axis in the thermodynamic limit,
consistent with the absence of a physical singularity for $y > y_c$.

From these results we are led to the conclusion that in fact
the locus of zeros in the thermodynamic limit {\it is not}
the unit circle, although due to the relatively small lattices
studied here we certainly do not offer this as a proof.
Rather, we believe the nature of the locus of zeros remains 
an open and interesting question.

\begin{table}
\caption{The distance from the origin and the imaginary 
part of the first two zeros of the three-state Potts model.
$Abs(x_1)$ and $Im(x_1)$ are the modulus and the imaginary part
of the first zero, $x_2$ is the second zero, 
and the last row is the BST extrapolation to infinite size.}
\begin{tabular}{rlllll}
$L$ &$Abs(x_1)$ &$Im(x_1)$ &$Abs(x_2)$ &$Im(x_2)$ \\
\hline
3  &1.133269811535 &0.525147232092 &1.154497346584 &1.064547354702 \\
4  &1.080426920767 &0.309148097981 &1.095611066859 &0.711951325609 \\
5  &1.054600270108 &0.205103734779 &1.065723514328 &0.488677034595 \\
6  &1.039822577595 &0.146911393618 &1.048300166208 &0.353845470678 \\
7  &1.030488924546 &0.110897277587 &1.037169667607 &0.268016265956 \\
8  &1.024179322221 &0.086960474252 &1.029587245456 &0.210307724775 \\
9  &1.019696144103 &0.070189130073 &1.024169532034 &0.169679308232 \\
10 &1.016386584326 &0.057951942531 &1.020153066105 &0.139981471036 \\
11 &1.013868175716 &0.048731329669 &1.017086499226 &0.117596130069 \\
12 &1.011903888317 &0.041599753769 &1.014688183571 &0.100288119772 \\
$\infty$ &1.0000(1) &0.00002(7) &1.0000(3) &0.000(1) \\
\end{tabular}
\end{table}

\begin{figure}
\epsfbox{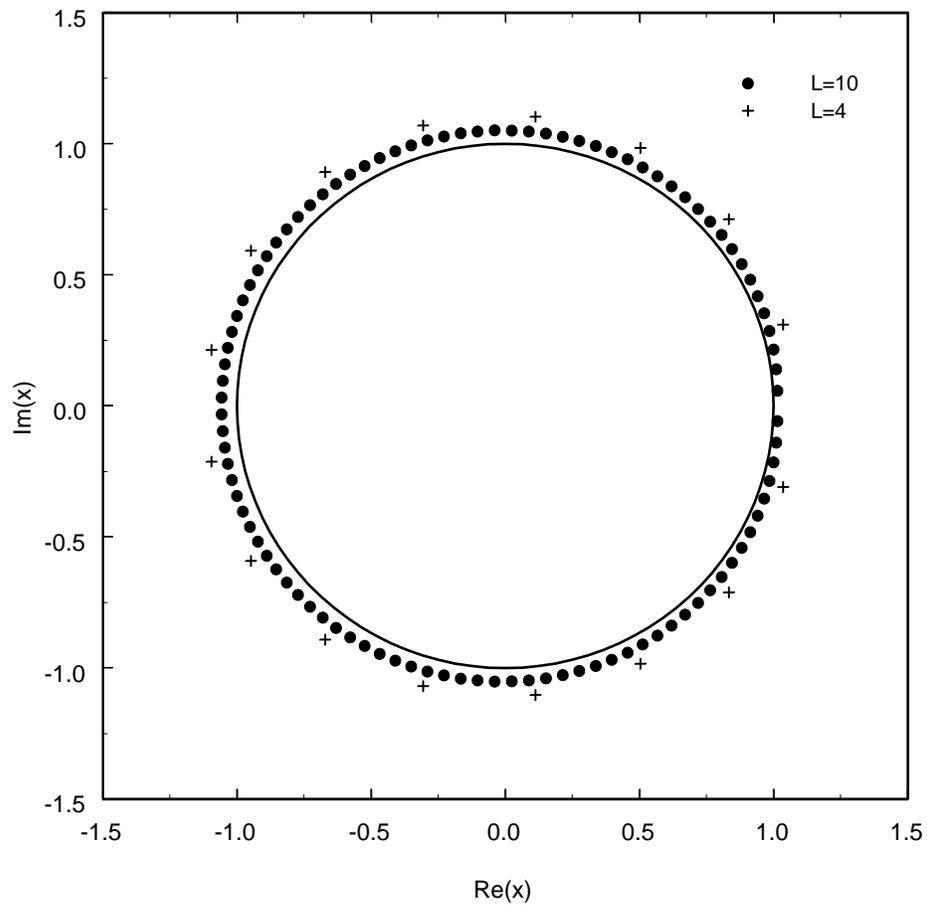}
\caption{Zeros of the three-state Potts model in the complex
$x$-plane at $y=y_c$ for $L=4$ and $L=10$ with cylindrical 
boundary conditions.}
\end{figure}

\begin{figure}
\epsfbox{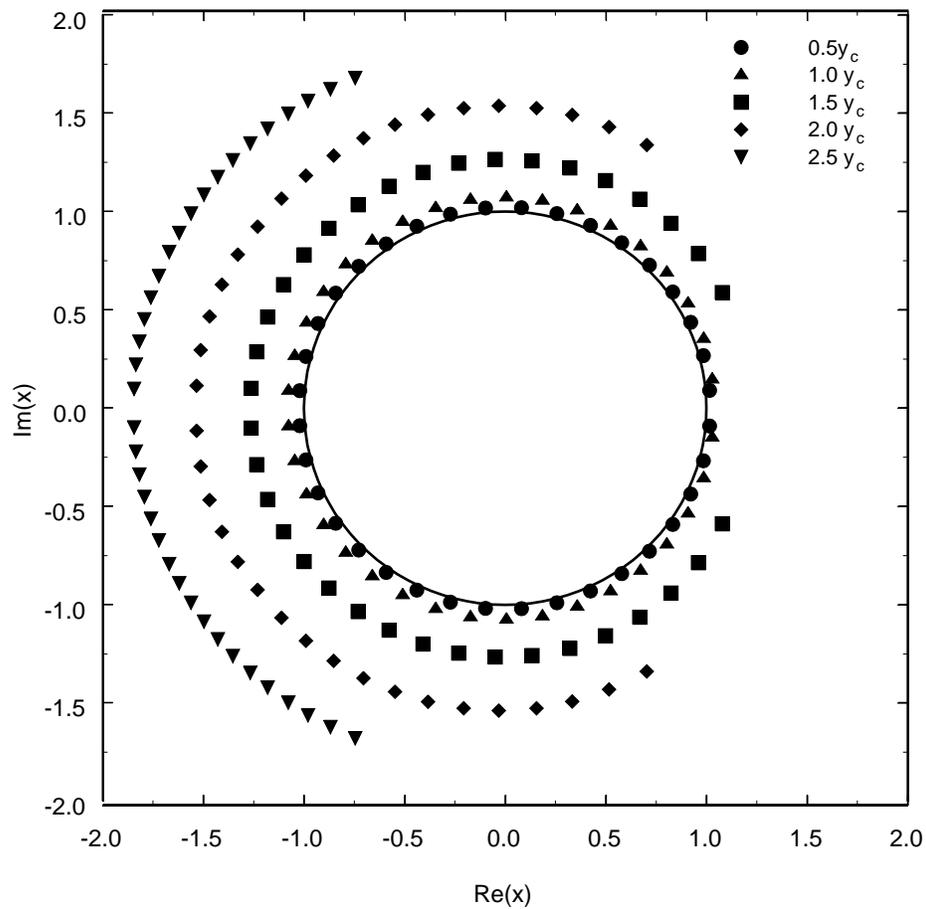}
\caption{Zeros of the three-state Potts model
in the complex $x$-plane for several values of $y$ 
($L=6$ and cylindrical boundary conditions).}
\end{figure}

\begin{figure}
\epsfbox{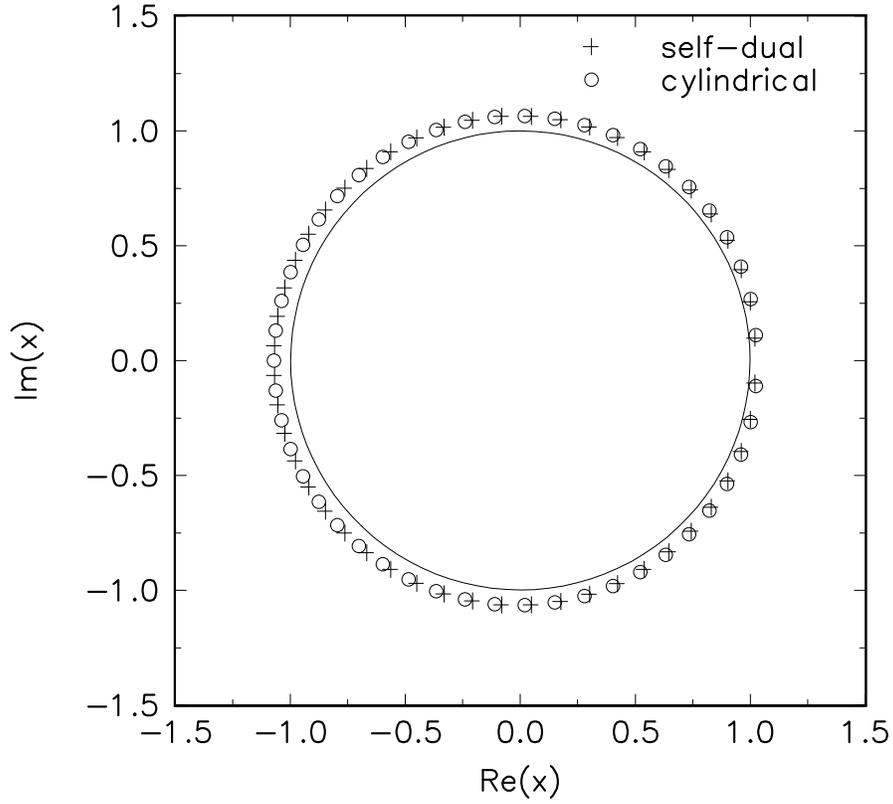}
\caption{Zeros of the three-state Potts model in the complex
$x$-plane at $y=y_c$ for $L=7$ with self-dual boundary conditions
(plus symbols) and cylindrical boundary conditions (open circles).}
\end{figure}

\begin{figure}
\epsfbox{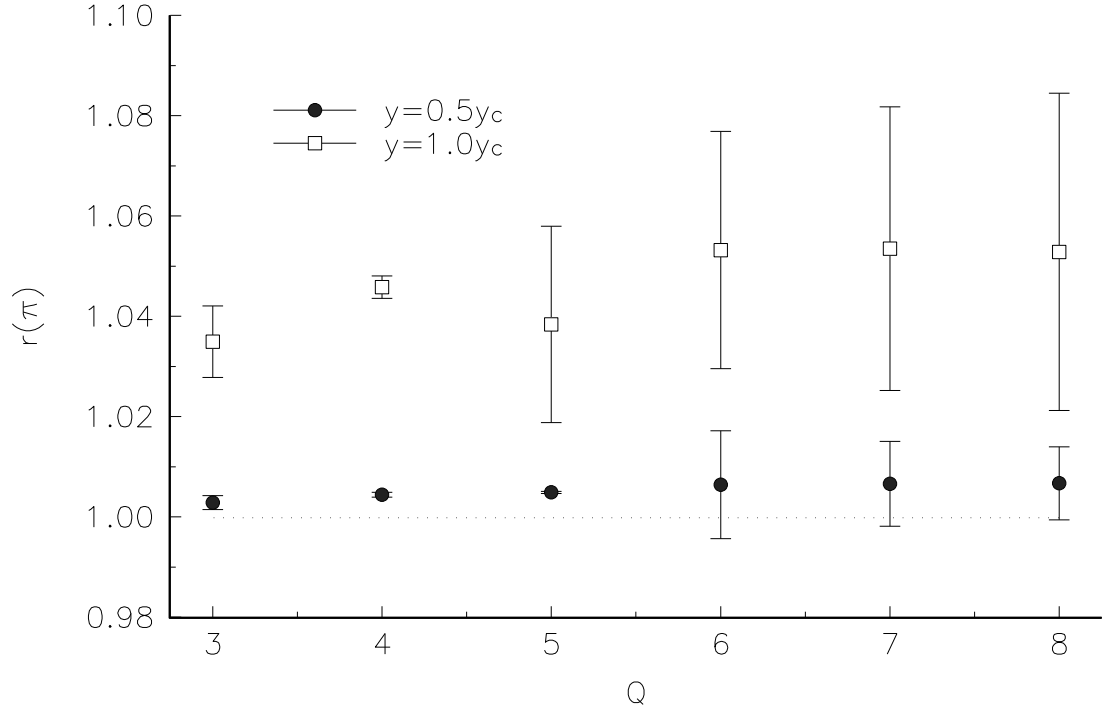}
\caption{Modulus of the zero at $\theta=\pi$
extrapolated to infinite size for $3\le Q\le8$ at 
$y=0.5y_c$ and $y=y_c$ with cylindrical boundary conditions.}
\end{figure}

\begin{figure}
\epsfbox{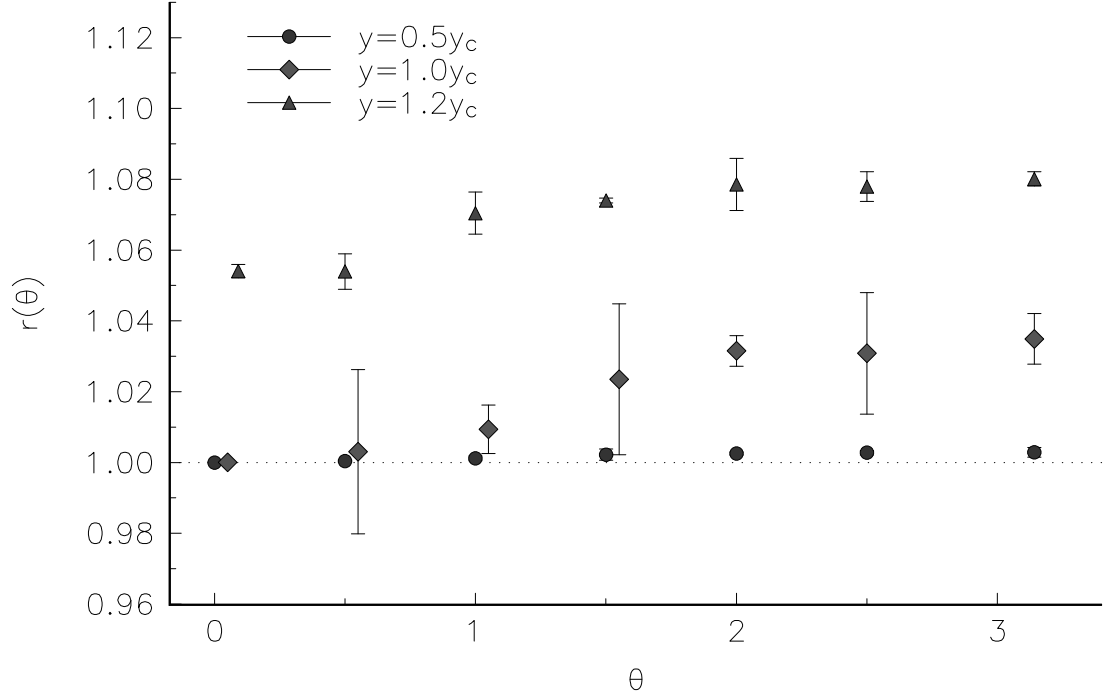}
\caption{Modulus of the locus of zeros as a function of angle for  
the three-state Potts model at $y=0.5y_c$, $y_c$, and $1.2y_c$
with cylindrical boundary conditions.
The slight horizontal off-set for data for $y=y_c$ is for clarity only.
However, the off-set of the edge singularity for $y=1.2y_c$ from
$\theta=0$ is real.}
\end{figure}

\end{document}